\documentclass{aa}
\usepackage{graphicx}
\usepackage{txfonts}

\newcommand{\diff}{\mbox{${\rm d}$}}

\newcommand{\ebv}{\mbox{$E_{B\!-\!V}$}}
\newcommand{\feh}{\mbox{\rm [{\rm Fe}/{\rm H}]}}

\newcommand{\Msun}{\mbox{$M_{\odot}$}}
\newcommand{\Teff}{\mbox{$T_{\rm eff}$}}

\newcommand{\comment}[1]{}
\newcommand{\beq}{\begin{equation}}
\newcommand{\eeq}{\end{equation}}
\newcommand{\beqa}{\begin{eqnarray}}
\newcommand{\eeqa}{\end{eqnarray}}
        \def\smallskip{\vskip 2pt}

\begin{document}

\title{Theoretical isochrones in several photometric systems} 
\subtitle{II. The Sloan Digital Sky Survey $ugriz$ system}

\author{L\'eo~Girardi\inst{1} \and 
 Eva K. Grebel\inst{2,3} \and 
 Michael Odenkirchen\inst{3} \and 
 Cesare Chiosi\inst{4}}
\institute{
 Osservatorio Astronomico di Trieste -- INAF, 
	Via G.B.\ Tiepolo 11, I-34131 Trieste, Italy \and 
 Astronomisches Institut der Universit\"at Basel, Venusstrasse 7,
        D-4102 Binningen, Switzerland \and
 Max-Planck-Institut f\"ur Astronomie, K\"onigstuhl 17, D-69117 Heidelberg,
        Germany \and
 Dipartimento di Astronomia, Universit\`a di Padova,
	Vicolo dell'Osservatorio 2, I-35122 Padova, Italy
}

\offprints{L\'eo Girardi \\ \email{Lgirardi@ts.astro.it}} 

\date{To appear in Astronomy \& Astrophysics}

\abstract{
Following Paper I, we provide extended tables of bolometric 
corrections, extinction coefficients,
stellar isochrones, and integrated magnitudes and 
colours of single-burst stellar populations, for the
Sloan Digital Sky Survey (SDSS) $ugriz$ photometric system.
They are tested on comparisons with DR1 data for a few 
stellar systems, namely the Palomar~5 and NGC~2419 globular
clusters and the Draco dSph galaxy.
}

\titlerunning{Isochrones in the SDSS system}
\authorrunning{L. Girardi et al.}
\maketitle

\section{Introduction}
\label{intro}

The Sloan Digital Sky Survey (SDSS; see York et al. 2000)
is one of the most impressive astronomical campaigns 
ever carried out. 
It aims at providing photometry and subsequent spectroscopy for objects
covering about one quarter of the entire sky.
SDSS photometry is being obtained near-simultaneously in five
broad-band filters in drift-scan mode, which 
results in highly homogeneous data.  The effective exposure
time for imaging is approximately 54\,s with a limiting magnitude
of $r\sim22.6$.  The primary science goals of the SDSS 
are at extragalactic and cosmological studies such as 
galaxy evolution as a function of redshift
(e.g., G\'omez et al.\ 2003; Eisenstein et al.\
2003; Blanton et al.\ 2001), the search for high-redshift
quasars (e.g., Fan et al.\ 2001, 2003),  weak lensing (e.g.,
Fischer et al.\ 2000; McKay et al.\ 2002), and the large-scale
structure distribution of galaxies (e.g., Dodelson et al.\ 2002;
Zehavi et al.\ 2002).  However, the SDSS also provides a huge
database for various areas of stellar and Galactic astronomy, e.g.,
the search for rare or special stellar objects (e.g., Margon et al.\
2002; Hawley et al.\ 2002; Helmi et al.\ 2003), 
the study of resolved star clusters and nearby Milky Way satellites
(e.g., Odenkirchen et al.\ 2001a,b), and studies of Milky Way
substructure (e.g., Chen et al. 2001; Newberg et al. 2002; Yanny
et al. 2003).  For an overview of SDSS science, see Grebel (2001).  
Our paper aims at providing tools for the interpretation of SDSS
stellar photometry data, i.e., isochrones transformed to the SDSS
photometric system.

The SDSS began regular operations in 2000 and will run until
2005.  The SDSS data products include images in five passbands, 
photometrically and astrometrically calibrated object catalogs 
with a wealth of information on the properties of the recorded 
objects, and wavelength- and flux-calibrated spectra with redshifts.
An Early Data Release (EDR) of SDSS commissioning data 
occurred in July 2001 (Stoughton et al.\ 2002),
and Data Release 1 (DR1) took place in May 2003 (DR1; Abazajian et al. 2003).
The currently publicly available SDSS photometry data 
cover 2099 square degrees, about one fifth 
of the total anticipated survey area.  The subsequent data 
releases are planned to occur on a yearly basis.  The 
final one is scheduled for 2006.  It is estimated that the 
SDSS will ultimately comprise some $8\cdot10^7$ stars with high-quality
five-band photometry, including Galactic halo and disk field stars,
stars in globular clusters, and in nearby resolved dwarf galaxies.

The SDSS has designed, defined, and calibrated its own
photometric system (Fukugita et al.\ 1996; Gunn et al.\ 1998;
Smith et al.\ 2003), 
which is characterized by the following particularities:
(1) the use of a modified Thuan-Gunn broadband filter system 
called $ugriz$;
(2) a zero-point definition in the AB magnitude system;
(3) the use of a modified, non-logarithmic, definition
of the magnitude scale (Lupton, Gunn, \& Szalay 1999).
As a consequence, SDSS photometry cannot be easily transformed into
traditional systems, or at least any transformation is likely to
lead to a significant loss of the photometric information
contained in SDSS data of fainter sources.  This presents a problem 
not only for users
of the SDSS databases, but also for astronomers who obtain new data
in the SDSS filters now offered at various observatories.

Therefore, in order to fully take advantage of the valuable
and growing SDSS database and to enable the quantitative 
interpretation of data obtained in the SDSS filter system elsewhere,
converting stellar models directly into the SDSS system 
is an obvious requirement. This is the goal of the present paper. 
This paper is a continuation of a series of papers dedicated 
to the conversion of theoretical stellar models to a wide variety of 
photometric systems.
In Paper~I (Girardi et al.\ 2002), we have described the 
assembly of a large library of stellar spectra, a simple
formalism to compute tables of bolometric corrections from this, 
and the application of these corrections to a large database of 
theoretical isochrones. The systems there considered were
the Johnson-Cousins-Glass, Washington, HST/WFPC2, HST/NICMOS,
and ESO Imaging Survey ones (for the WFI, EMMI, and SOFI cameras
in use at the European Southern Observatory at La Silla, Chile). 

The basic procedures and the input stellar data involved in 
the present work are the same as in Paper~I. 
Thus, here we will skip most of the description,
detailing just the particular aspects that are inherent to 
the SDSS photometric system.  In Section 2, we describe the
SDSS AB and asinh magnitude systems and the synthetic photometry
required to transform theoretical
isochrones into these systems.  In Section 3,
we apply these isochrones to multi-color DR1 data.  
Section 4 contains our conclusions.

\section{Synthetic photometry and results}
\label{sec_synphot}

The basic procedures of synthetic photometry are extensively discussed
in Paper~I, as well as the adopted library of synthetic and empirical 
spectra. Skipping all the details, we just recall that our primary target 
is the derivation of bolometric corrections $BC_{S_\lambda}$, for each
filter $S_\lambda$ and for each star of intrinsic spectrum $F_\lambda$.
According to section 2.2 of Paper~I, they are given by
	\beqa
BC_{S_\lambda} & = & M_{\rm bol, \odot} 
	- 2.5\,\log \left[ 
		4\pi (10\,{\rm pc})^2 F_{\rm bol}/L_\odot
		\right] \label{eq_bcfinal} 
	\\ \nonumber
	&& + 2.5\,\log\left(
	\frac { \int_{\lambda_1}^{\lambda_2} 
		\lambda F_\lambda \, 10^{-0.4A_\lambda} 
		S_\lambda \diff\lambda }
		{ \int_{\lambda_1}^{\lambda_2} 
		\lambda f^0_\lambda S_\lambda \diff\lambda } 
	\right)
	- m_{S_\lambda}^0
	\eeqa
where $f^0_\lambda$ is a reference spectra which produces the
standard magnitudes $m_{S_\lambda}^0$.

We have initially constructed a library of intrinsic stellar spectra
$F_\lambda$, consisting basically of:
spectral library:
	\begin{itemize}
\item ATLAS9 non-overshooting models (Castelli et al. 1997; 
Bessell et al. 1998), complemented with
\item blackbody spectra for $\Teff>50\,000$ K, 
\item Fluks et al. (1994) empirical M-giant spectra, extended 
with synthetic ones in the IR and UV, and modified shortward of
4000 \AA so as to produce reasonable \Teff--$U-B$ and \Teff--$B-V$ 
relations for cool giants,
\item Allard et al. (2000; see also Chabrier et al. 2000) "DUSTY99" 
synthetic spectra for M, L and T dwarfs.
	\end{itemize}

Again, we refer to Paper I for all
definitions and details. We now simply shift to a description of
the SDSS filter system and derived isochrones.

	\begin{figure*}
	\resizebox{\hsize}{!}{\includegraphics{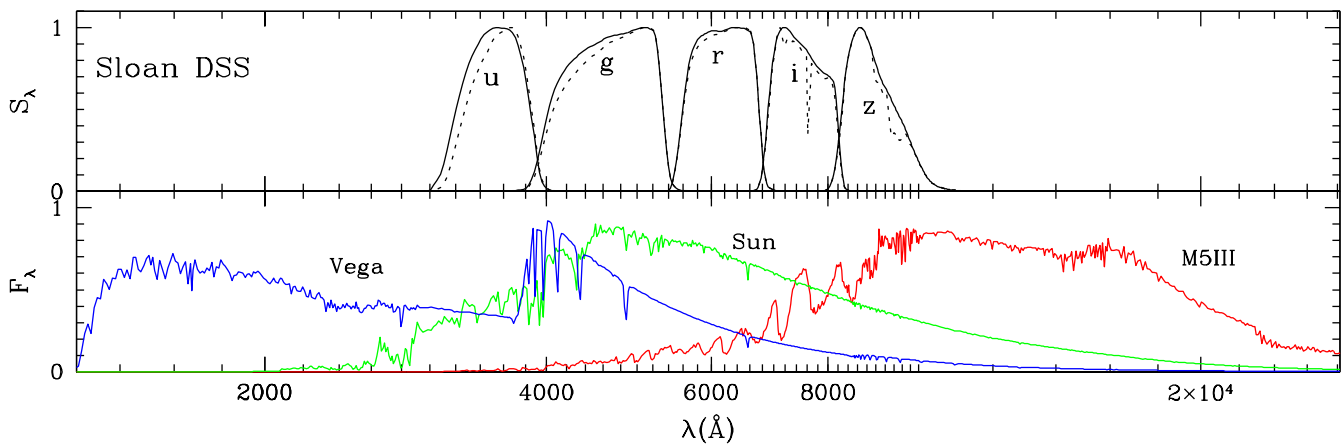}}
	\caption{The SDSS
filter+detector transmission curves $S_\lambda$ adopted in this
work. They refer to the filter and detector throughputs
as seen through airmasses of 1.3 (dashed lines) 
at Apache Point Observatory. For the sake of comparison,
the curves for a null airmass (solid lines) are also presented.
All curves are re-normalized to their maximum value of $S_\lambda$.
The bottom panel presents 
the spectra of Vega (A0V), the Sun (G2V), and a M5 giant,
in arbitrary scales of $F_\lambda$. The $\lambda$ scale here adopted
is the same as in Figure 3 of Paper~I.}
	\label{fig_filtri}
	\end{figure*}

The SDSS photometric system is defined by Fukugita et al.\ 
(1996); in this paper, it is described by means of normal
AB magnitudes. Later on, Lupton et al. (1999) devised a modified way 
of expressing magnitudes -- namely a change from a logarithmic to a
inverse hyperbolic sine function of flux -- that has been adopted by 
the SDSS project. It is worth discussing both cases separately. 

\subsection{Isochrones in the usual AB magnitude scale}

As described by Fukugita et al.\ (1996), the SDSS uses an 
AB magnitude system (Oke \& Gunn 1983), which considerably simplifies the 
derivation of bolometric correction tables:
The zero-points are completely defined by using a 
synthetic spectrum of constant flux per unit frequency, without 
making any reference to the spectrum of real stars. This is
detailed in Section 2.3.2 of Paper~I.
Here, suffice it to recall that 
the adopted reference spectra and magnitudes are
$f_{{\rm AB},\nu}^0=3.631\times10^{-20}\,\,\, {\rm erg \, 
s^{-1} \, cm^{-2} \, Hz^{-1}}$, and
$m_{{\rm AB}, \nu}^0=0$ for all filters, respectively.

Therefore, the bolometric corrections and
AB magnitudes of the SDSS system have been computed in the
same way used in Paper~I to generate HST/WFPC2 and HST/NICMOS 
AB magnitudes. 

For the transmission curves $S_\lambda$ of the $ugriz$ filters, 
we use the SDSS filter curves available from the DR1 webpages\footnote{
The URL for the SDSS filter curves is:\\
{\tt  www.sdss.org/dr1/instruments/imager/index.html\#filters}}.
They are illustrated in Fig.~\ref{fig_filtri}, and 
include the filter and detector throughputs, and the atmosphere
as seen through an airmass of 1.3 at Apache Point Observatory.
Details of these transmission curves may 
change as future measurements of the filters are done. 

It is worth mentioning that Fukugita et al.\ (1996) use
the spectra of real stars -- namely the Oke \& Gunn (1983) 
spectrophotometric standards -- in order to define the 
zero-points of SDSS photometry, even suggesting a revision 
in the spectra of these spectrophotometric standards. 
But actually these revisions regard
only the transformation of observations into 
the standard ABmag system, and do not imply any change in
the definition of AB magnitudes. Therefore, the zero-points of
our theoretical models are not affected: they are, 
by construction, strictly in the ABmag system. 

\subsection{Conversion to the Lupton et al.\ modified magnitude
scale}

Lupton et al.\ (1999) suggested a modified magnitude
definition where the classical logarithmic scale 
	\beq
m(f) = m_0 - 2.5\log f
	\label{eq_magclassic}
	\eeq
(where $m$ is a magnitude, $m_0$ is a zero-point, 
and $f$ is the photon flux as integrated over a filter 
pass-band) is replaced by an inverse hyperbolic sine function
	\beq
\mu(f) = (m_0-2.5\log b') - a \sinh^{-1} (f/2b') \,\,\,
	\label{eq_maglupton}
	\eeq
where $a=2.5\log e$, and $b'$ is the constant (in photon flux units) 
that gives $\mu(0)=m_0 - 2.5 \log b'$ for a null flux.
In practice, $b'$ is related to the limiting magnitude of a given 
photometric survey, and has to be furnished together with the
apparent $\mu$ in any of its data releases.
This magnitude definition reproduces the traditional definition for
objects measured with a signal-to-noise $> 5$, avoids problems
with negative fluxes for very faint objects, and retains a well-behaved
error distribution for fluxes approaching zero.  Hence it is primarily
of importance for objects near the detection limit. 

It is clear that this definition of magnitude is not compatible 
with the formalism we adopt to derive bolometric corrections. 
Actually, {\em basic quantities like bolometric corrections, 
absolute magnitudes, and distance modulus, 
cannot be defined in any simple way if we use the 
Lupton et al. scale}, because it is a non-logarithmic one. 
As a corollary, we can say that such a scale represents a 
convenient way to express apparent magnitudes and colours 
near the survey limit
(as demonstrated by Lupton et al.\ 1999), but represents a
complication if we want to represent absolute magnitudes.

Considering this, we do not even try to express our theoretical
models by means of Lupton et al.\ (1999) modified magnitude scale.
We do, however, provide a prescription of how to convert 
absolute magnitudes $M_{S_\lambda}$ -- 
given by our models in the AB system -- 
into an apparent $\mu_{S_\lambda}$ -- as given in the SDSS data 
releases. This can be done in the following way:
	\begin{enumerate}
\item convert from absolute to apparent magnitudes using
the usual definitions of distance modulus and absorption, i.e.,
$m_{S_\lambda} = M_{S_\lambda ,0} + (m-M)_0 + A_{S_\lambda}$;
\item convert from classical apparent magnitudes to a
photon flux, i.e.\ $f=\int(\lambda/hc)\diff\lambda$ in the 
case of ABmags; this requires knowledge of the effective 
throughputs in each pass-band $S_\lambda$, referring to the 
complete instrumental configuration (pratical hints on 
this step, regarding SDSS DR1 data, can be found in the URL
{\tt http://www.sdss.org/dr1/algorithms/fluxcal.html});
\item convert the photon flux to Lupton et al.\ (1999) modified 
magnitude scale by means of Eq.~\ref{eq_maglupton}, using the $b'$ 
constant typical of the observational campaign under consideration.
	\end{enumerate}

Of course, the procedure is not as simple as one would like.
Since at good signal-to-noise ratios ($f>b'$) the Lupton et al.\ 
scale coincides with the classical definition of magnitudes, 
the question arises whether it is necessary at all to convert
models to Lupton et al.\ (1999) scale. In fact, for most analyses of 
stellar data it will not be worthwhile, since one is rarely tempted to 
derive astropysical quantities from stars measured with large
photometric errors. 

For diffuse and faint objects like distant galaxies, however, the 
situation might well be the opposite one: even relatively noisy data
may contain precious astrophysical information. Useful hints about
the dominant stellar populations may result, for instance, from
a comparison between the integrated 
magnitudes and colours of single-burst stellar populations 
(provided in this paper in the usual magnitude scale) 
to those of faint galaxy structures from SDSS (given in the 
Lupton et al.\ scale). If this is the case, the conversion problem 
has to be faced.

\subsection{Extinction coefficients}
\label{sec_alambda}

\begin{table*}
\caption{Computed $A_\lambda/A_V$ values for a few stars and extinction cases.}
\label{tab_alambda}
\begin{tabular}{lll|ll|lllll}
	\hline\noalign{\smallskip}
\multicolumn{3}{l|}{Star:} & \multicolumn{2}{l|}{extinction:} & 
	\multicolumn{2}{l}{$A_\lambda/A_V$ for:} \\
\Teff & $\log g$ & $[{\rm M/H}]$ & $A_V$ & $R_V$ & 
 $u$ & $g$ & $r$ & $i$ & $z$ \\
	\noalign{\smallskip}\hline\noalign{\smallskip}
5777 & 4.44 & 0 & 0.5 & 3.1 & 1.574 & 1.199 & 0.879 & 0.675 & 0.490 \\
\noalign{\smallskip}\hline\noalign{\smallskip}
5777 & 4.44 & 0 & 1.0 & 3.1 & 1.574 & 1.196 & 0.878 & 0.674 & 0.490 \\
5777 & 4.44 & 0 & 2.0 & 3.1 & 1.573 & 1.190 & 0.876 & 0.672 & 0.489 \\
5777 & 4.44 & 0 & 3.0 & 3.1 & 1.571 & 1.184 & 0.875 & 0.671 & 0.488 \\
5777 & 4.44 & 0 & 1.0 & 5.0 & 1.331 & 1.126 & 0.903 & 0.732 & 0.564 \\
\noalign{\smallskip}\hline\noalign{\smallskip}
3500 & 0.00 & 0 & 0.5 & 3.1 & 1.551 & 1.154 & 0.874 & 0.662 & 0.481 \\
3500 & 2.00 & 0 & 0.5 & 3.1 & 1.551 & 1.175 & 0.877 & 0.660 & 0.479 \\
3500 & 4.50 & 0 & 0.5 & 3.1 & 1.558 & 1.178 & 0.874 & 0.663 & 0.482 \\
3750 & 0.00 & 0 & 0.5 & 3.1 & 1.553 & 1.154 & 0.872 & 0.667 & 0.484 \\
3750 & 2.00 & 0 & 0.5 & 3.1 & 1.553 & 1.170 & 0.874 & 0.665 & 0.484 \\
3750 & 4.50 & 0 & 0.5 & 3.1 & 1.558 & 1.178 & 0.874 & 0.665 & 0.485 \\
4000 & 0.00 & 0 & 0.5 & 3.1 & 1.558 & 1.156 & 0.872 & 0.670 & 0.485 \\
4000 & 2.00 & 0 & 0.5 & 3.1 & 1.557 & 1.170 & 0.873 & 0.669 & 0.486 \\
4000 & 4.50 & 0 & 0.5 & 3.1 & 1.560 & 1.180 & 0.874 & 0.668 & 0.486 \\
4500 & 0.00 & 0 & 0.5 & 3.1 & 1.567 & 1.166 & 0.875 & 0.672 & 0.487 \\
4500 & 2.00 & 0 & 0.5 & 3.1 & 1.569 & 1.175 & 0.875 & 0.672 & 0.487 \\
4500 & 4.50 & 0 & 0.5 & 3.1 & 1.568 & 1.185 & 0.875 & 0.672 & 0.488 \\
5000 & 0.00 & 0 & 0.5 & 3.1 & 1.564 & 1.179 & 0.876 & 0.673 & 0.489 \\
5000 & 2.00 & 0 & 0.5 & 3.1 & 1.570 & 1.184 & 0.877 & 0.673 & 0.489 \\
5000 & 4.50 & 0 & 0.5 & 3.1 & 1.574 & 1.189 & 0.877 & 0.673 & 0.489 \\
6000 & 2.00 & 0 & 0.5 & 3.1 & 1.565 & 1.204 & 0.879 & 0.675 & 0.490 \\
6000 & 4.50 & 0 & 0.5 & 3.1 & 1.574 & 1.202 & 0.879 & 0.675 & 0.490 \\
8000 & 4.50 & 0 & 0.5 & 3.1 & 1.573 & 1.221 & 0.883 & 0.678 & 0.491 \\
10000 & 4.50 & 0 & 0.5 & 3.1 & 1.573 & 1.231 & 0.885 & 0.679 & 0.491 \\
12500 & 4.50 & 0 & 0.5 & 3.1 & 1.578 & 1.235 & 0.886 & 0.680 & 0.492 \\
15000 & 4.50 & 0 & 0.5 & 3.1 & 1.581 & 1.237 & 0.886 & 0.680 & 0.493 \\
20000 & 4.50 & 0 & 0.5 & 3.1 & 1.585 & 1.240 & 0.886 & 0.680 & 0.494 \\
30000 & 4.50 & 0 & 0.5 & 3.1 & 1.588 & 1.243 & 0.887 & 0.681 & 0.495 \\
40000 & 4.50 & 0 & 0.5 & 3.1 & 1.589 & 1.245 & 0.887 & 0.681 & 0.495 \\
50000 & 5.00 & 0 & 0.5 & 3.1 & 1.590 & 1.246 & 0.887 & 0.681 & 0.495 \\
4000 & 2.00 & -2 & 0.5 & 3.1 & 1.560 & 1.173 & 0.873 & 0.670 & 0.487 \\
4000 & 4.50 & -2 & 0.5 & 3.1 & 1.564 & 1.179 & 0.872 & 0.670 & 0.487 \\
6000 & 4.50 & -2 & 0.5 & 3.1 & 1.576 & 1.209 & 0.879 & 0.675 & 0.490 \\
10000 & 4.50 & -2 & 0.5 & 3.1 & 1.573 & 1.231 & 0.885 & 0.679 & 0.491 \\
\noalign{\smallskip}\hline\noalign{\smallskip}
\multicolumn{5}{l|}{Schlegel et al. (1998):} & 
 1.579 & 1.161 & 0.843 & 0.639 & 0.453 \\
\multicolumn{5}{l|}{Fiorucci \& Munari (2003), $R_V=3.1$:} & 
 $1.61^{1.58}_{1.69}$ & $1.19^{1.16}_{1.20}$ & $0.83^{0.84}_{0.85}$ & 
	$0.61^{0.61}_{0.64}$ & $0.45^{0.45}_{0.48}$ \\
	\noalign{\smallskip}\hline
\end{tabular}
\\
\begin{tabular}{lll|ll|llllllll}
	\hline\noalign{\smallskip}
\Teff & $\log g$ & $[{\rm M/H}]$ & $A_V$ & $R_V$ & 
 $U$ & $B$ & $V$ & $R$ & $I$ & $J$ & $H$ & $K$ \\
	\noalign{\smallskip}\hline\noalign{\smallskip}
5777 & 4.44 & 0 & 
0.5 & 3.1 & 1.562 & 1.298 & 1.006 & 0.826 & 0.600 & 0.292 & 0.186 & 0.116 \\
\noalign{\smallskip}\hline\noalign{\smallskip}
\multicolumn{5}{l|}{Schlegel et al. (1998)} & 
   &  &  &  &  &  &  &  \\
\multicolumn{5}{l|}{(for CTIO and UKIRT filters):} & 
 1.521 & 1.324 & 0.992 & 0.807 & 0.601 & 0.276 & 0.176 & 0.112 \\
\multicolumn{5}{l|}{Fiorucci \& Munari (2003), $R_V=3.1$:} & 
 $1.58^{1.54}_{1.62}$ & $1.31^{1.25}_{1.32}$ & $0.99^{1.00}_{1.00}$ & 
 $0.77^{0.75}_{0.79}$ & $0.56^{0.56}_{0.58}$ & $0.27^{0.27}_{0.28}$ &
 $0.17^{0.17}_{0.18}$ & $0.11^{0.11}_{0.12}$ \\
	\noalign{\smallskip}\hline
\end{tabular}
\end{table*} 

The basic formalism of synthetic photometry as introduced in 
Paper~I, allows an easy assessment of 
the effect of interstellar extinction on the output data.
As can be readily seen in Eq.~(\ref{eq_bcfinal}), 
each stellar spectrum $F_\lambda$ in our database 
can be reddened by applying a given extinction curve $A_\lambda$, 
and hence the bolometric corrections computed as usual. 
The difference between the BCs derived from reddened spectra 
and the original (unreddened) ones, divided by the amount of 
total reddening in a reference passband (say $A_V)$, 
gives the so-called extinction coefficients
	\beq
A_\lambda/A_V = \frac{BC_{S_\lambda}(A_V) - BC_{S_\lambda}(0)} {A_V} 
	\eeq
These quantities are of course useful for consistently computing 
the amount of extinction expected in different passbands.

In order to derive $A_\lambda/A_V$ values, we first assume the
Cardelli et al. (1989) extinction curve for a typical Galactic 
total-to-selective ratio of $R_V=A_V/E_{B-V}=3.1$.
The derived values of $A_\lambda/A_V$ are tabulated in 
Table~\ref{tab_alambda}, for a few different stellar spectra
(from ATLAS9 models) ranging in \Teff, $\log g$, 
and metallicity $[{\rm M/H}]$. 

The first line in the table presents our reference
$A_\lambda/A_V$ values, namely those derived for a Sun-like 
yellow dwarf ($\Teff=5777$~K, $\log g=4.44$, $[{\rm M/H}]=0$;
Kurucz 1993) with low reddening and a normal $R_V=3.1$ extinction
curve. They will be adopted in the simple comparisons with data 
to follow in Sect.~\ref{sec_tests}. 
For the same star and for the sake of comparison, 
$A_\lambda/A_V$ coefficients
are presented also for the Johnson-Cousins-Glass system
as defined by Bessell \& Brett (1988) and Bessell (1990).

Then, the table presents a sequence of $A_\lambda/A_V$ ratios 
for increasing values of total extinction $A_V$ at constant $R_V=3.1$.
It illustrates that $A_\lambda$ has a very small, 
non-linear trend with $A_V$ (see also Grebel \& Roberts 1995).
The $R_V=5.0$ value, suitable for dense star-forming regions as the 
Orion Nebulae, is also considered. In this case, $A_\lambda/A_V$ ratios
become very different from the previous ones.

Most of Table~\ref{tab_alambda} is dedicated to present $A_\lambda/A_V$ 
ratios obtained for stars of different \Teff\ and $\log g$,
for both solar and metal poor compositions ($[{\rm M/H}]=0$ and $-2$,
respectively), in the limit of low total extinction and for
$R_V=3.1$. They are all derived from Bessell et al. (1997) ATLAS9 
spectra. 
It can be noticed that $A_\lambda/A_V$ depends in a non-negligible
way on \Teff\ and $\log g$, and to a much lower extent
on $[{\rm M/H}]$. All these effects are thoroughly discussed
in Grebel \& Roberts (1995). For the present paper, suffice it to 
mention that we are able to provide $A_\lambda/A_V$ values
for any other extinction curve and stellar spectra, upon request.

We also note that the extinction coefficients we find for 
the SDSS system are somewhat different from those mentioned in the 
SDSS Early Data Release by Stoughton et al. (2002),
and derived by Schlegel et al. (1998; see values in 
Table~\ref{tab_alambda}).
Assessing the detailed reasons for these differences is beyond 
the scope of this paper. Anyway, we remark that 
Schlegel et al. (1998) compute their extinction values 
using a very different spectrum (an elliptical galaxy one),
which is likely one of the main sources of the discrepancy.
On the other hand, our $A_\lambda/A_V$ values are also
different from those computed by Fiorucci \& Munari (2003) 
using essentially the same stellar spectra and 
extinction curves. In this latter case, 
the discrepancies are likely to be ascribed to the different
numerical methods and, to a lower extent, to their use of 
Fukugita et al.~(1996) filter transmission curves instead 
of DR1 ones. Among the possible differences in numerical methods,
it is remarkable that both Schlegel et al. (1998) and
Fiorucci \& Munari (2003) use the concept of photon-energy 
integration in their definition of $A_\lambda$, instead of 
the photon-count integration we perform in
ours Eq.~\protect\ref{eq_bcfinal} (see Paper~I for a 
discussion on this topic). 

\subsection{Presently available isochrones}
\label{sec_isoc}

All isochrone sets from the Padova group have
been converted into the SDSS ABmag system. Their input stellar
tracks and particularities are listed in 
Table~1 of Paper~I. Here, suffice it to recall the main isochrone
sets available:
\begin{itemize}
\item A ``basic set'' derived by using the Girardi et al. (2000)
evolutionary tracks for $0.15 \le M/\Msun \le 7$ together
with previous Padova tracks for massive stars of $M\ge 7\Msun$ 
(from Bressan et al.\ 1993; Fagotto et al.\ 1994ab; 
Girardi et al.\ 1996). This set has also been complemented with
newest tracks (Girardi et al. 2003) for massive $Z=0.001$ stars,
and covers a very wide range of metallicities (from $Z=0.0001$
to $0.03$).
It also represents a sort of fusion between the
Bertelli et al. (1994) isochrones for ages $t\la10^8$ yr, 
with the Girardi et al.\ (2000) ones for $t>10^8$ yr; 
\item A set for metal-free ($Z=0$) stars, 
from Marigo et al. (2001).
\item The Salasnich et al. (2000) isochrones for both solar-scaled
and $\alpha$-enhanced chemical compositions, and moderate to high
metallicity range (from $Z=0.008$ to $0.07$).
\item The Marigo \& Girardi (2001) isochrones, derived by complementing
the Girardi et al. (2000) tracks with the very detailed TP-AGB
sequences from Marigo et al. (1999), 
available for $Z=0.004$, $0.008$ and $0.019$.
\end{itemize}

In the following sections, we will only use the above-mentioned
``basic set'' of isochrones, which covers the widest range 
of ages and metallicities. 

The list of isochrone sets is in continuous expansion and
subject to revision at any time (see for instance the main 
addition anticipated by Marigo et al. 2003). 
Importantly, {\em the present paper
does not present any new isochrone set, but just their
transformation into a new system. Users of SDSS isochrones
should better refer also to the original source of stellar data},
listed in the items above and detailed in the electronic 
database that contains all data tables.

	\begin{figure*}
	\resizebox{\hsize}{!}{\includegraphics{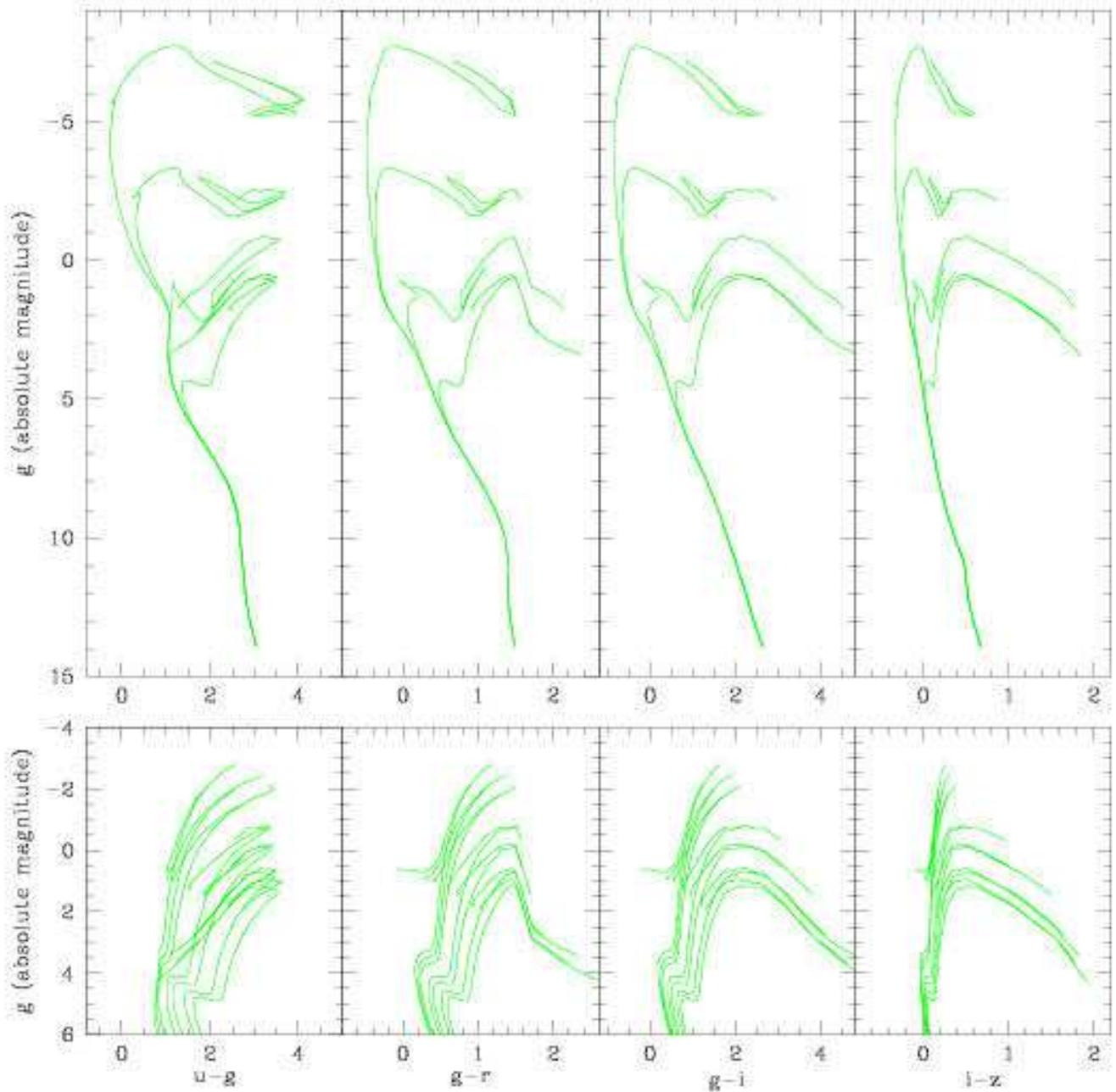}}
	\caption{Isochrones in several CMDs of SDSS photometry. 
Top panels: a sequence of solar-metallicity ($Z=0.019$) isochrones
of increasing ages, namely $10^7$, $10^8$, $10^9$, $10^{10}$ yr, 
going from the brightest to faintest turn-off.
Bottom panels: a sequence of 12-Gyr old isochrones of increasing
metallicities, namely $Z=0.0001$, $0.0004$, $0.001$, $0.004$, $0.008$, 
$0.019$ and $0.03$, going from the bluest to the reddest turn-off.}
	\label{fig_isoc}
	\end{figure*}


Figure~\ref{fig_isoc} shows two different isochrone sequences --
taken from the ``basic set'' -- in CMDs of SDSS photometry:
one of varying age at solar metallicity, and another of varying
metallicity at old ages. The figure well illustrates the expected
locus of some main evolutionary phases (main sequence, RGB, red clump,
HB, cepheid loop, AGB, red supergiants), as well as the expected 
appearance of CMDs for star clusters, in this yet unfamiliar 
photometric system.

\section{Tests using DR1 data}
\label{sec_tests}

We chose three different objects in order to test the isochrones:
Two globular clusters (Palomar 5 and NGC 2419) and one nearby dwarf
spheroidal galaxy (Draco), which are all three included in the DR1
database.  In fact, these three objects are the {\em only}\ globular clusters
and nearby resolved dwarf galaxies currently available from the public
SDSS database.  All three objects have been extensively studied in the
past in standard photometric systems such as Johnson-Cousins and also
have spectroscopic metallicity measurements.  We can use them now for
isochrone consistency checks for old
populations at intermediate ($-1.4$ dex) and low ($\sim -2$ dex)
metallicities.  We require that Johnson and SDSS data can be fit with
isochrones in these two photometric systems that use the same 
parameters (internal consistency), and that these parameters are 
consistent with previous results from the literature (external
consistency).

Before proceeding, we recall that a preliminary version of our
isochrones -- using 2001 filter tranmission curves from
Strauss \& Gunn, as retrieved from 
{\tt http:\-//archive\-.stsci.\-edu/sdss/documents/response.\-dat} -- 
have been recently compared to data for the open cluster M\,48 by 
Rider et al. (2004). Their data were obtained using $u'g'r'i'z'$ filters
available at the CTIO Curtis-Schmidt, US Naval Observatory (USNO) 1.0m,
and SDSS 0.5m Photometric telescopes.
Thus, Rider et al.'s work shall be considered as an additional test 
to the theoretical isochrones we are presenting in this paper.

\subsection{Palomar 5}

Palomar 5 is a sparse halo globular cluster at a distance of $\sim 23$ kpc
(Harris 1996). Its current mass is only 4.5 to $6\cdot 10^3$ M$_{\odot}$
(Odenkirchen et al.\ 2002), and it is undergoing significant tidal
disruption as evidenced by its two well-defined, symmetric tidal tails
(Odenkirchen et al.\ 2001a, 2003; Rockosi et al.\ 2002).  With a metallicity 
of $\feh\sim -1.4$ dex (Harris 1996), it is the most metal-rich resolved 
single-age stellar population presently available from the public SDSS 
photometry database.  Therefore we will use this cluster despite its 
sparseness for our isochrone comparison.

	\begin{figure}
	\resizebox{\hsize}{!}{\includegraphics{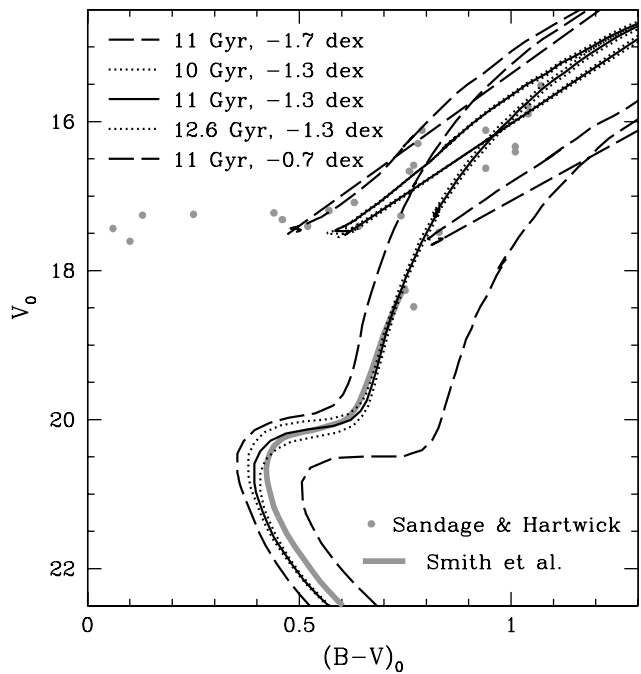}}
	\caption{Reddening-corrected
         color-magnitude diagram of Palomar 5 in the Johnson
         $B$ and $V$ filters.  The data points (grey dots) were taken from 
         Sandage \& Hartwick (1977).  The thick grey solid line represents 
         the mean
         location of the lower red giant branch, subgiant branch, and 
         main sequence as given by Smith et al.\ (1986).  The thinner
         black lines -- dashed, dotted, and solid --
         are Johnson isochrones from Girardi et al. (2002)
         as detailed in the legend (sequence from left to right).
         The $Z=0.001$ ($-1.3$ dex) isochrone with an age of 11 Gyr
         provides the closest match to the Smith et al. fiducial.  Older
         and younger isochrones (dotted lines) have main-sequence
         turn-offs that are too faint or too luminous, respectively.}
	\label{Fig_Pal5_BV}
	\end{figure}

We first attempt to constrain the age and other parameters of
Pal\,5 using Johnson photometry
data that extend well below the main-sequence turn-off.
If the Johnson and the SDSS isochrones are internally consistent, then
the parameters that we derive from the Johnson data should also apply
to the SDSS data.
Published Johnson $B$, $V$ photometry is available from Sandage
\& Hartwick (1977) and Smith et al.\ (1986).  In Fig.\ \ref{Fig_Pal5_BV},
we show data points from Sandage \& Hartwick's photoelectric study and the
fiducial sequence derived by Smith et al., together with isochrones from
Girardi et al. (2002) in the generic Johnson system.  We find the 
$Z=0.001$ (or $\feh\sim -1.3$ dex) and $\log(t)=10.05$ (11 Gyr) 
isochrone to provide the closest fit to the overall slope of the fiducial
and to the location of the main-sequence turn-off.  This metallicity is 
in good agreement with the literature value (see above).
We also found that the use
of a slightly higher reddening than found in earlier studies (cf.\
Sandage \& Hartwick 1977), $\ebv = 0.04$ mag, improves the fit. 
The deviations between the fiducial main sequence and the
isochrone along the color axis may be caused, in part, by differences 
between the filter
and detector combination used by Smith et al. and the Johnson
filter definition used by Girardi et al., and by the procedure used by Smith
et al. (1986) to derive the mean fiducial values.  For instance, Smith
et al. do not find evidence for a binary main sequence and consequently do
not correct for it, while new deep CCD data (Koch et al. 2003, in prep.) show
clear evidence for the presence of binaries.   

The SDSS photometric database provides an estimate of the Galactic foreground
extinction along the line of sight of each catalogued object, which is based
on the extinction maps of Schlegel, Finkbeiner, \& Davis (1998) and which 
thus applies primarily to objects outside of the Milky Way.  The mean
reddening value resulting from the Schlegel et al.\ (1998) maps is 
$\langle\ebv\rangle = 0.056$ mag.  We choose to
adopt the same constant reddening value of $\ebv = 0.04$ mag as used 
above for the Johnson photometry and convert it  
into the SDSS system using the extinction coefficients
given in Table~\ref{tab_alambda}.   As an investigation into 
large-scale extinction properties in the region of Pal 5 has shown, the 
reddening {\em at the location of the cluster} is low and homogeneous (see 
Odenkirchen et al.\ 2003, Figure 12), which further supports our use of
one single, fixed reddening value.
The amount of interstellar extinction also depends on  
the target star's effective temperature, surface gravity, and metallicity
as detailed in Grebel \& Roberts (1995), and as illustrated 
for SDSS filters in Table~\ref{tab_alambda}. 
In the limit of low extinction (for say $\ebv\la0.1$), 
the magnitude of these effects is small,
so that we do not take them into account in this section.

        \begin{figure}
        \resizebox{\hsize}{!}{\includegraphics{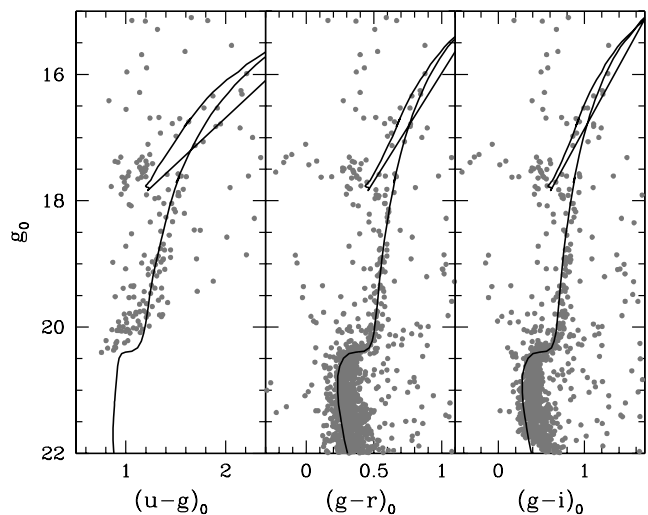}}
        \caption{Color-magnitude diagrams of Palomar 5 in different 
         SDSS filters.  The grey dots represent stars with photometric 
         uncertainties $<0.1$ mag.
         The solid line is one of our new SDSS isochrones
         with an age of $\log(t)=10.05$ and $Z=0.001$, i.e., with the best-fit
         parameters as
         found in Fig.\ \ref{Fig_Pal5_BV}.  This isochrone provides
         the closest match to the SDSS data.  
         }
        \label{Fig_Pal5_SDSS}
        \end{figure}

Fig.~\ref{Fig_Pal5_SDSS} shows the resulting, de-reddened 
CMDs of Pal 5 using different SDSS filter combinations.  Only stars
whose photometric uncertainties are below 0.1 mag are plotted.  Colors
including the SDSS $z$-band were omitted, since no well-defined giant branch
was visible for Pal 5.  Our SDSS
isochrones were shifted by the appropriate distance modulus (see above).  
As before, an isochrone with a metallicity of $Z=0.001$ and an age of
11 Gyr ($\log t = 10.05$)
approximates the observed CMD best, especially in the part 
corresponding to the RGB.
A careful inspection of the figure however reveals that below the
turn-off level this isochrone runs slightly to the blue 
($\la0.1$~mag) with respect to the mean colour of 
main sequence stars. Such a discrepancy 
could be caused either by a real colour shift in the models, 
or by the presence of binaries in the data. 

\subsection{NGC 2419}

        \begin{figure}
        \resizebox{\hsize}{!}{\includegraphics{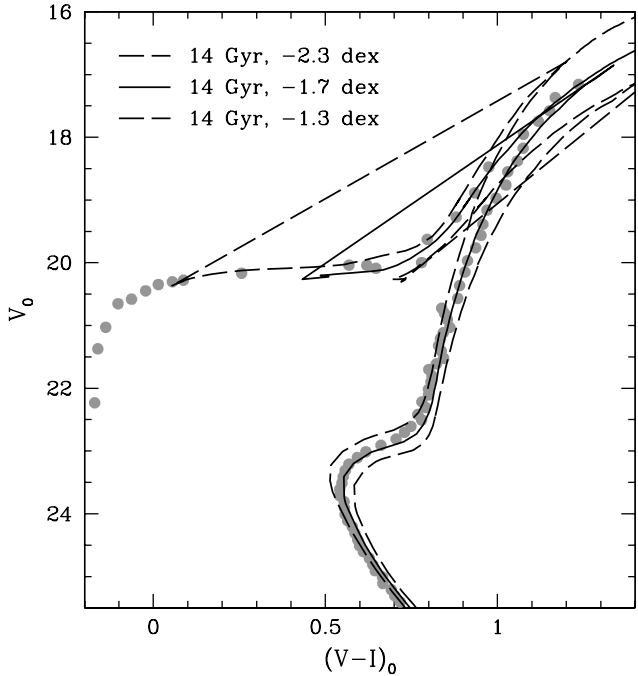}}
        \caption{$V$, $I$ color-magnitude diagram of NGC 2419.
         The grey dots mark the fiducial derived for this cluster by
         Harris et al. (1997).  Also shown are three of our new  
         SDSS isochrones
         with an age of $\log(t)=10.10$.  Of these, the $Z=0.0004$ isochrone
         ($\feh=-1.7$ dex; solid line)
         provides the closest match to the data and best reproduces the
         slope of the red giant branch when using the reddening and 
         distance modulus found by Harris et al. (1997).
         }
        \label{Fig_N2419_VI_CMD}
        \end{figure}

NGC 2419 is a massive, very distant outer halo globular cluster.
Harris et al.\ (1997) quote a distance of $\sim 81$ kpc and a
metallicity of $\feh=-2.14$ dex.
The $V$, $I$ data of Harris et al. extend well below the main-sequence 
turn-off of NGC~2419 near $V\sim 24$ mag.  Since the SDSS limiting 
magnitude is much brighter than this, we use the fiducial points provided 
by Harris et al. to find the best-fitting Padua isochrone.  For a  
reddening of $\ebv = 0.11$ mag as suggested by Harris et al. (1997), we find 
the closest fit for an
isochrone with $\log(t) = 10.15$ ($\sim 14$ Gyr) and $Z=0.0004$ 
($\feh\sim -1.7$ dex).
The $Z=0.0001$ isochrone (Fig.\ \ref{Fig_N2419_VI_CMD}), while being closest
to the spectroscopic metallicity, is too steep and does 
not reproduce the slope of the red giant branch.  This appears to be a
general property of metal-poor Padua isochrones (see, e.g., Fig.\ 5 in
Grebel 1999) and does not affect the goals of our present work, since 
we care primarily about establishing an intrinsically consistent set
of isochrones in different photometric systems.  

        \begin{figure}
        \resizebox{\hsize}{!}{\includegraphics{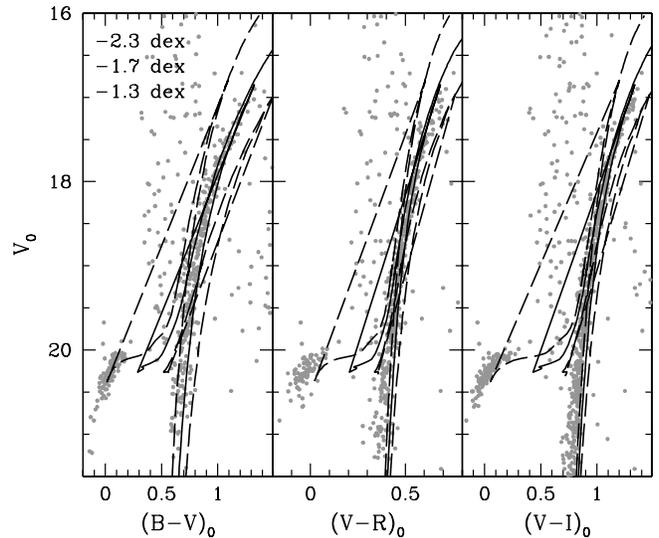}}
        \caption{Johnson/Kron-Cousins $BVRI$ color-magnitude diagrams
         of NGC 2419 based on Stetson's (2000) homogeneous photometry
         (grey dots).  Superimposed we show Johnson/Kron-Cousins isochrones
         from Girardi et al. (2000) with the same parameters as in 
         Fig.\ \ref{Fig_N2419_VI_CMD}.  As before, the $-1.7$ dex, 14
         Gyr isochrone matches the multi-color data well, 
	apart from a general shift to redder colours and 
	fainter magnitudes. 
         }
        \label{Fig_N2419_BVRI_CMD}
        \end{figure}

We note that Harris et al.\ (1997) found NGC\,2419 to be essentially
indistinguishable in age from the ancient, metal-poor halo globular cluster
M92, provided that these two clusters do not show pronounced differences
in their chemical abundance ratios, particularly in their $\alpha$ elements.
In this context it is interesting to note that Newberg et al. (2003)
recently found NGC\,2419 to be probably associated with a distant tidal
arm of the Sagittarius dwarf spheroidal galaxy (see also Zhao 1998).  While 
this could be 
taken as an indication that this cluster's $\alpha$ element abundances
differ from those measured in the Galactic halo, Shetrone, C\^ot\'e,
and Sargent (2001) found the element ratios in NGC\,2419 very similar to
those in M92 (based on the analysis of one star).  Hence we cautiously
consider it justified to use isochrones with
$\alpha$ element ratios akin to those in the Galactic halo.

The $Z=0.0004$, 14 Gyr
isochrone with the above reddening and distance also provides a 
good fit to CMDs of NGC\,2419 in other Johnson/Kron-Cousins
bands as shown in Fig.\ \ref{Fig_N2419_BVRI_CMD}. 
There seems to be a general shift of the isochrones to redder 
colours and fainter magnitudes, which might be indicating an 
overestimation of the reddening towards this cluster.
Such a shift was not apparent in the former 
Fig.\ \ref{Fig_N2419_VI_CMD}, hence suggesting an off-set in the
photometric data rather than in the models.
The input photometry in this case was
taken from Stetson's (2000) on-line database of homogeneous $BVRI$ 
photometry for star clusters and resolved galaxies, which includes only
photometry based on multiple independent measurements.

        \begin{figure}
        \resizebox{\hsize}{!}{\includegraphics{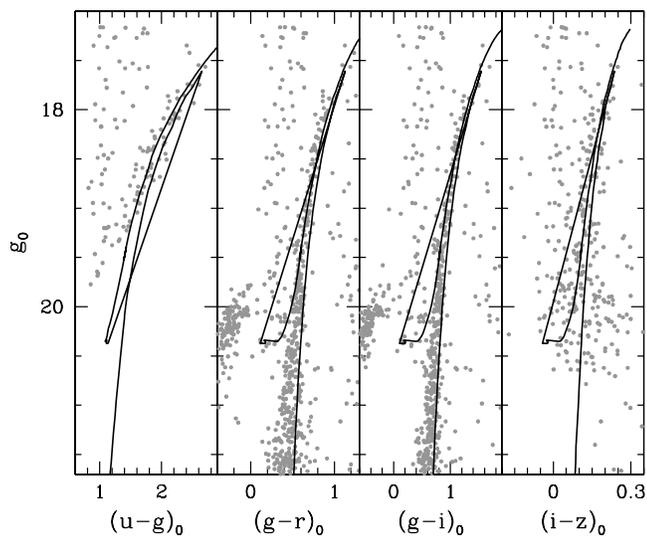}}
        \caption{
         SDSS {\em ugriz} color-magnitude diagrams
         of NGC 2419 for stars with photometric uncertainties $<0.1$ mag
         (grey dots).  Superimposed we show our new SDSS isochrone
         with the parameters found to work best for Johnson/Kron-Cousins
         photometry (see Fig. \ref{Fig_N2419_VI_CMD} and 
         \ref{Fig_N2419_BVRI_CMD}).  This isochrone also provides a
         fully consistent fit to the multi-color SDSS data.
         }
        \label{Fig_N2419_ugriz_CMD}
        \end{figure}

In Fig.\ \ref{Fig_N2419_ugriz_CMD} the SDSS $ugriz$ photometry is shown
together with the newly transformed isochrone that also provided the
best fit to the Johnson/Kron-Cousins data.  As before for the 
Johnson/Kron-Cousins photometry, we used a fixed reddening of $\ebv
= 0.11$ mag, which we transformed using the SDSS extinction coefficients
listed in Table 1. As in the case of Johnson/Kron-Cousins photometry,
there is a good match between the SDSS isochrone and
the SDSS data but for a slight overall shift of the isochrones to 
redder colours and fainter magnitudes. The magnitude of this shift 
in the different passbands indicates that a better match to the
NGC\,2419 data could be obtained with smaller values of reddening,
closer to $\ebv=0$ than to the value of $\ebv=0.11$ mag here 
considered.
If we assume this is the case, the good overall fit of
isochrones to NGC\,2419 data shows that we get consistent 
results also for metal-poor abundances. 
%

\subsection{Draco}

While the dwarf spheroidal galaxy Draco is dominated by old populations, 
it differs from single-age, single-population objects like globular
clusters in showing a considerable abundance spread of more than one dex
in [Fe/H] (Lehnert et al.\ 1992; Shetrone, Bolte, \& Stetson 1998; 
Shetrone et al. 2001; and references therein).  Draco's mean spectroscopic 
metallicity is 
$\sim -1.9$ dex according to Lehnert et al. (1992; based on 14 red giants) 
and $\sim -2.0$ dex
following Shetrone et al.\ (2001; based on six giants).
Aparicio et al. (2001) found a mean photometric metallicity of $-1.8$
dex when applying Padua isochrones to Johnson $B,R$ photometry.  Bellazzini
et al.\ (2002) compared $V$, $I$ photometry of Draco's red giants to
globular cluster fiducials and derived a mean photometric metallicity
of $-1.7$ dex.  Aparicio et al. (2001) did not see indications
of a metallicity gradient across the galaxy within the $\sim 1$ deg$^2$
area they studied.  On the other hand, Klessen, Grebel, \&
Harbeck (2003), who analyzed a much larger area based on SDSS data,
noted a population gradient in the sense that blue horizontal branch stars
show a spatially more extended distribution
than red horizontal branch stars.  Spatial gradients in the horizontal
branch morphology are commonly found in dwarf spheroidal galaxies
(see Hurley-Keller, Mateo, \& Grebel 1999; Grebel 2000;
Harbeck et al.\ 2001; Grebel, Gallagher, \& Harbeck 2003) and
are likely either due to age spreads in the central regions of these 
galaxies, due to metallicity spreads, or a combination thereof.

        \begin{figure}
        \resizebox{\hsize}{!}{\includegraphics{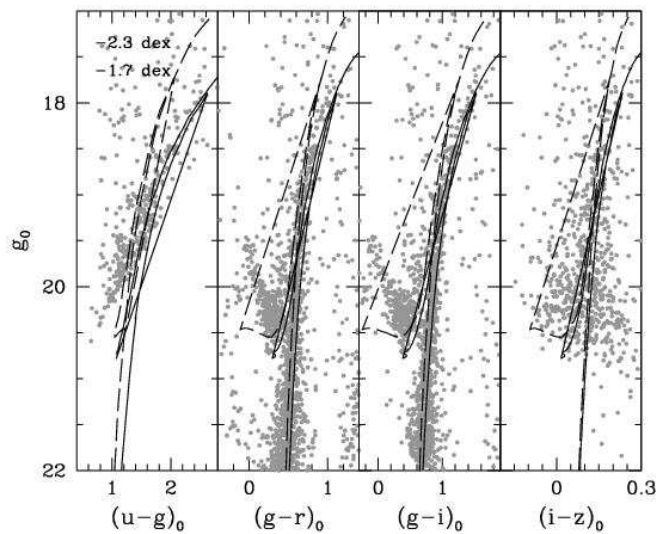}}
        \caption{
         SDSS {\em ugriz} color-magnitude diagrams
         of the central $9'\times9'$ of the Draco dwarf spheroidal galaxy
         for stars with photometric uncertainties $<0.1$ mag
         (grey dots).  Superimposed we show our new SDSS isochrones
         for ages of $\log(t)=10.1$ and the parameters discussed in the text. 
         The main concentration of the red giant branch of Draco lies 
         blueward of the $-1.7$ dex isochrone (solid line).
         }
        \label{Fig_Draco_ugriz_CMD}
        \end{figure}

In order to test the new SDSS isochrones, we confine ourselves to the
central $9'\times9'$ region of Draco, which has the added advantage of 
reducing the contribution of contaminating Galactic foreground stars.  
As described above, we are very
likely sampling a mix of (predominantly old) populations with a range of
metallicities and an age spread.  The analysis by Aparicio et al.\ (2001),
using Padua isochrones, does indeed suggest a range of ages.  We therefore
choose an age of $\log(t) =10.1$ for the SDSS isochrones, which corresponds 
roughly to the mean age indicated by the star formation histories of
Aparicio et al.\ (2001). We adopt the Draco distance obtained 
through various methods by Bellazzini et al.\ (2002), i.e., $\sim 93$ kpc. 
The interstellar extinction along the line of sight to Draco is low but
variable (see the extinction map in Odenkirchen et al.\ 2001b, their 
Fig.\ 1).  We used both the individual, star-by-star
Schlegel et al.\ (1998) extinction values available
from the DR1 database and a constant value of $\ebv = 0.03$ mag together
with the extinction coefficients in Table 1 and did not find qualitative
differences in the resulting CMDs for the central region considered here.  

The resulting SDSS CMDs are shown in Figure \ref{Fig_Draco_ugriz_CMD},
using the above parameters and individual dereddening.  The $Z=0.0004$
(or $\feh=-1.7$ dex is closest to the main body of the red giant 
branch of Draco in the different filters and provides a red envelope.  
If we were to derive a mean metallicity from the comparison with the
SDSS isochrones, it would be closer to the lower metallicity estimates
($-1.8$ to $-1.9$ dex).  As pointed out by Shetrone et al.\ (2001), 
Draco's [$\alpha$/Fe] ratio is lower by $\sim 0.2$ dex
than what is found in the Galactic
halo, which will affect age estimates when isochrones with Galactic halo
$\alpha$ ratios are used, but this should not affect the internal
consistency check for the SDSS isochrones presented here.  Considering
the photometric uncertainties and the known metallicity spread in Draco,
we consider the comparison between the isochrones at fixed 
metallicity spacing with the multi-color
SDSS data to demonstrate very satisfactory consistency.

\section{Concluding remarks}
\label{sec_conclu}

We have computed bolometric corrections and extinction coefficients
specific for the SDSS ABmag photometric system. They have been
applied to transform previous Padova isochrones and integrated
magnitudes of single-burst stellar populations into SDSS $ugriz$ 
absolute magnitudes. 
All data tables mentioned in this paper are available in the
web site {\tt http://pleiadi.pd.astro.it}. They have a structure
similar to those already released with Paper~I, and extended
descriptions are also included in the form of electronic files.

Comparisons between the present isochrone sets and SDSS DR1
data, together with the recent results by Rider et al. (2004), 
are encouraging.
Our hope is that this database will be useful for the analysis
of the huge amount of photometric data that has been, 
and will yet be, released by the SDSS project. Obvious 
applications go from the derivation of parameters for
star clusters and nearby dwarf galaxies (distances, reddenings, 
ages and metallicities), to the isolation of particular 
objects in colour-colour diagrams, to the application in 
analyses of star counts and Galactic structure. 
The further use of SDSS filters in other observatories,
in programs not related to the original survey,
may well expand the possible range of use for these tables.

\begin{acknowledgements}
L.G.\ thanks funding by the COFIN 2002028935\_003, and the 
hospitality by MPIA during a visit.
E.K.G. acknowledges partial funding through
the Swiss National Science Foundation. 

Funding for the creation and distribution of the SDSS Archive has been 
provided by the Alfred P. Sloan Foundation, the Participating Institutions, 
the National Aeronautics and Space Administration, the National Science 
Foundation, the U.S.\ Department of Energy, the Japanese Monbukagakusho, 
and the Max Planck Society. The SDSS Web site is {\tt http://www.sdss.org/}. 

The SDSS is managed by the Astrophysical Research Consortium (ARC) for the
Participating Institutions. The Participating Institutions are The University 
of Chicago, Fermilab, the Institute for Advanced Study, the Japan 
Participation Group, The Johns Hopkins University, Los Alamos National 
Laboratory, the Max-Planck Institute for Astronomy (MPIA), the Max-Planck 
Institute for Astrophysics (MPA), New Mexico State University, University 
of Pittsburgh, Princeton University, the United States Naval Observatory, 
and the University of Washington.

This research has made use of NASA's Astrophysics Data System.

\end{acknowledgements}


\end{document}